\documentclass[doublecol]{epl2} 
\usepackage{graphicx}
\usepackage{dcolumn}
\usepackage{bm}
\usepackage{amsmath}

\title{Granular Impact Model as an Energy-Depth Relation}
\shorttitle{Granular Impact Model as an Energy-Depth Relation} 

\author{A. H. Clark \and R. P. Behringer}
\shortauthor{A. Clark \etal}

\institute{Department of Physics, Duke University - Box 90305,
Durham, NC 27708, USA}


\abstract{{Velocity-squared drag forces are common in
    describing an object moving through a granular material. The
    resulting force law is a nonlinear differential equation, and
    closed-form solutions of the dynamics are typically obtained by
    making simplifying assumptions. Here, we consider a generalized
    version of such a force law which has been used in many studies of
    granular impact. We show that recasting the force law into an
    equation for the kinetic energy versus depth, $K(z)$, yields a
    \emph{linear} differential equation, and thus general closed-form
    solutions for the velocity versus depth. This approach also has
    several advantages in fitting such models to experimental data,
    which we demonstrate by applying it to data from 2D impact
    experiments. We also present new experimental results for this
    model, including shape and depth dependence of the
    velocity-squared drag force.}}

\begin{document}

\maketitle

\section{Introduction}
A dense granular material which is struck by a high-speed object
exerts a decelerating force on the intruder, the nature of which is
important for many applications, such as soil penetration
tests\cite{Campbell1986,Dexter2007}, meteor impacts\cite{Guttler2012},
and ballistic
applications\cite{Nesterenko1996,Forrestal1992}. Additionally,
understanding the dynamics of intruder motion in a granular material,
as well as the granular flow around it, is a fundamental problem in
granular physics. To probe this process, it is common to drop or push
an intruder into a granular material and record the dynamics and/or
the force, $F$, exerted on the intruder. In general, the dynamics 
depend on the microscale physical characteristics of the grains and
intruder, and may show large fluctuations, {as in \cite{Clark2012}}.
 
A common approach
\cite{Euler1745,Poncelet1829,Allen1957,Tsimring2005,Katsuragi2007,Goldman2008,Goldman2010,Takehara2010,Clark2012}
dating back to the times of Euler and Poncelet is to use space-time
averaged macroscopic force laws, where the various terms in the law
are empirical expressions based on assumptions of relevant physical
principles. These terms are typically assumed to depend on the depth,
$z$, as well as on the intruder velocity, $v$, taken to be strictly
vertical, i.e. $v=\dot{z}$. {Most of these models have the following generalized form}:
\begin{equation}
F=m\ddot{z}=mg-f(z)-h(z)\dot{z}^2,
\label{eqn:forcelaw}
\end{equation}
where dots denote time derivatives. This force law contains three
terms: {gravity; a depth-dependent static term, $f(z)$, often
  taken to be linear in $z$; and a collisional term proportional to
  the square of the intruder speed, $h(z)\dot{z}^2$, where $h(z)$ is
  often assumed to be constant. Here, $z$ is measured from the
  original unperturbed surface, with $z=0$ as the point of initial
  contact. These force laws are intended as coarse-grained
  descriptions of local granular processes, similar in spirit to the
  static and collisional terms in a general coarse-graining
  description, such as that formulated by Goldenberg and
  Goldhirsch\cite{Goldenberg2006}.  Regardless of the justification,
  they are often quite successful in describing the average dynamics
  of the intruder trajectories. }

{ However, theoretical or experimental application of these
  models reveals several difficulties. First, eq.~\eqref{eqn:forcelaw}
  cannot be integrated to obtain the closed-form solutions for the
  trajectory, $z(t)$ and $v(t)$. To obtain the final stopping time and
  depth, one must assume specific forms for $f(z)$ and $h(z)$,
  typically that they are constants \cite{Allen1957, Forrestal1992,
    Goldman2008, Tsimring2005}.  But, this assumption is inconsistent
  with several experimental studies
  \cite{Katsuragi2007,Goldman2008,Goldman2010}. Additionally,
  experimental comparison to these models requires depth, velocity,
  and acceleration data for the intruder. Typically, this is done by
  examining data from trajectories with many different initial
  velocities\cite{Katsuragi2007,Goldman2008,Goldman2010,Takehara2010,Clark2012}. For
  a specific depth, $z=\zeta$, eq.~\eqref{eqn:forcelaw} becomes:
\begin{equation}
F=m\ddot{z}=mg-f(\zeta)-h(\zeta)\dot{z}^2.
\label{eqn:forcelawconstdepth}
\end{equation}
If this model is valid, plotting $\ddot{z}$ versus $\dot{z}^2$, calculated at $z=\zeta$, yields an approximately straight line of slope $h(\zeta)$ and intercept $g-f(\zeta)/m$. However, it is often difficult to obtain accurate acceleration data at short time scales, since accelerometers have limited time-resolution, and discrete differentiation from position data greatly amplifies measurement noise. Finally, acceleration measurements often contain large fluctuations at short time scales \cite{Forrestal1992,Goldman2008,Goldman2010,Clark2012}. Recent work \cite{Clark2012} has shown that these fluctuations are a physical aspect of the dynamics, connected to acoustic activity beneath the intruder. These large fluctuations make precise determination of $f(\zeta)$ and $h(\zeta)$ difficult, as shown in fig.~\ref{fig:accvsvsqr}. In contrast, the fluctuations in velocity data are considerably smaller, which is a crucial point in the analysis presented here.
\begin{figure}
\onefigure[scale=0.9,trim=0mm 0mm 0mm 0mm]{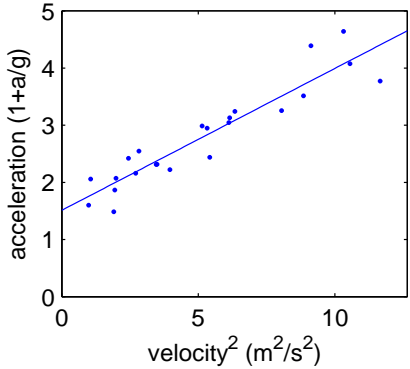}
\caption{{Plot of acceleration versus velocity squared for a circular intruder at a fixed depth, $z=\zeta$, where each of the 23 data points shown represents one trajectory from experiments described later in the text. The linear fit shown, $1+a/g=C_0+C_1v^2$, has $C_0=1.5$ and $C_1=0.25$, which is the best fit from a linear regression. However, due to the large acceleration fluctuations, the 95\% confidence intervals for $C_0$ and $C_1$, respectively, are (1.25,1.78) and (0.20,0.30). Thus, without significantly more trajectories, it is difficult to extract a precise value for $C_0$ or $C_1$, which correspond to $f(\zeta)$ and $h(\zeta)$, unless one makes some assumptions about $f(z)$ and $h(z)$, such as $h(z)=b=$~constant where $b$ is the average $h(z)$ for all depths.}}
\label{fig:accvsvsqr}
\end{figure}}

In this letter, we {demonstrate a new approach to applying this force-law model, where
eq.~\eqref{eqn:forcelaw}, a nonlinear differential equation for force versus time,
is reformulated into a \emph{linear} differential equation in kinetic energy. We show that this approach addresses all of the aforementioned issues. Mathematically, it allows formal closed-form trajectory solutions without specific assumptions on $f(z)$ and $h(z)$. These solutions then provide a natural way to experimentally measure $f(z)$ and $h(z)$ using only velocity and depth data, with no assumptions about the functional form of these terms. Using high-speed video data from two-dimensional impact experiments with bronze intruders and photoelastic disks, we study the dynamics using this new approach. This yields several important results, including a significantly higher collisional force at the point of impact for circular intruders, an effect which vanishes for intruders with elongated noses.}

\section{Kinetic Energy Formulation} 
We recast eq.~\eqref{eqn:forcelaw} from second order in time for $z$
to first order in depth for kinetic energy of the intruder,
$K=\frac{1}{2}m\dot{z}^2$, by using a relation that is familiar from
the work-energy theorem of mechanics: $m\ddot{z}=dK/dz$.
\begin{equation}
\frac{dK}{dz} = mg-f(z)-\frac{2h(z)}{m}K.
\label{eqn:KEmodel}
\end{equation}
This yields an inhomogeneous \textit{linear} ordinary differential
equation with (potentially) nonconstant coefficients, by contrast to
the nonlinear equation of motion, eq.~\eqref{eqn:forcelaw}, for
$\ddot{z}(t)$. The fact that eq.~\eqref{eqn:KEmodel} is a linear ODE
means that standard ODE methods immediately yield formal solutions for
$K(z)$:
\begin{equation}
K(z) = K_p(z)(K_0 + \phi (z)).
\label{eqn:KEsol2a}
\end{equation}
where $K_0$ is the kinetic energy at impact, 
\begin{equation} K_p(z) = \exp\left(-\int_0^z (2/m)h(z') dz'\right),
\end{equation}
and 
\begin{equation}
\phi = \int_0^{z}dz'[mg -f(z')]/K_p(z').
\end{equation}

This reduces the problem for the trajectory to a quadrature. The
velocity can be written as:
\begin{equation}
\dot{z}(z) =\frac{dz}{dt}=\left[\frac{2}{m} K_p(z)(K_0 + \phi (z))\right]^{1/2},
\label{eqn:vsol}
\end{equation}
and $z(t)$ follows by integrating and inverting:
\begin{equation}
t(z) =\int_0^z dz' \left[\frac{2}{m} K_p(z')(K_0 + \phi (z'))\right]^{-1/2}.
\label{eqn:trajsol}
\end{equation}

If the forms of $f(z)$ and $h(z)$ are simple, much of the calculation
of these integrals can by done explicitly. For example, using the commonly assumed forms $f(z)=f_0+kz$ and $h(z)=b$, we obtain
$K(z)$ as
\begin{equation}
K(z) = (K_0 -c_1)\exp(-c_2 z) + c_1 - c_3 z.
\label{eqn:k-z-simp}
\end{equation} 
Here, the constants are $c_1 = [(mg-f_0)c_2 + k]/{c_2}^2$, $c_2 =
2b/m$, and $c_3 = k/c_2 = km/(2b)$.

Even without integrating eq.~\eqref{eqn:trajsol}), it is possible to
find the stopping distance by setting $K(z_{stop}) = 0$, or
$\phi(z_{stop}) = -K_0$, yielding the stopping depth as a function of
impact energy, $K_0$. Specifically, for the common case described by
eq.~\eqref{eqn:k-z-simp}, the stopping depth, $z_{stop}$ satisfies
\begin{equation}
z_{stop}=\frac{m}{2b}\log\left[\frac{\frac{2b}{m} K_0+\left(f_0+\frac{km}{2b}\right)-mg}{\left(f_0+kz_{stop}+\frac{km}{2b}\right)-mg}\right]
\label{eqn:z-stop-simp}
\end{equation}
Note that if we take $f(z)$ as roughly constant, $f(z)=f_0$ and $k=0$, then
$z_{stop}$ increases logarithmically with $K_0$, as in \cite{Tsimring2005, Goldman2008, Forrestal1992}.
\begin{equation}
z_{stop}=\frac{m}{2b}\log\left[1+\frac{2b}{m}\left(\frac{K_0}{f_0-mg}\right)\right]
\label{eqn:highKEsol}
\end{equation}
This approximation is relevant in the limit of high impact energy,
where $b\dot{z}^2$ dominates. In the low energy limit, it predicts
$z_{stop}(K_0=0)=0$.  However, an intruder must be at least partially
submerged to be supported by frictional grains. This occurs at a depth
which increases with intruder size. So, when $K_0\rightarrow 0$, we
expect $z_{stop}(K_0=0)\sim D$. Note that eqs.~\eqref{eqn:k-z-simp}
and \eqref{eqn:z-stop-simp} may yield a nonzero upward force,
$F=\frac{dK}{dz}$, as the intruder comes to a stop, and we return
later to this issue.

\subsection{Measuring $f(z)$ and $h(z)$}
This formulation also provides a way to \textit{measure} $f(z)$ and
$h(z)$ in terms of experimental data for $z(t)$ and $\dot{z}(t)$,
without specific assumptions for $f(z)$ and $h(z)$.  Subtracting two
different trajectories, with different $K_0$'s (not necessarily
close), $K_i(z)=K_p(K_{i0}+\phi)$ and $K_j(z)=K_p(K_{j0}+\phi)$,
yields:
\begin{equation}
\frac{K_i(z)-K_j(z)}{K_{i0}-K_{j0}}\equiv K_p(z)=e^{-\int_0^z\frac{2h(z')dz'}{m}},
\label{eqn:ijdiff}
\end{equation}
which gives,
\begin{equation}
h(z)=-\frac{d}{dz}\left[\frac{m}{2}\log K_p(z)\right].
\label{eqn:hsol}
\end{equation}
To avoid numerical differentiation, we also use $\int h(z) dz$ in our
discussion below.  

Since the kinetic energy goes to zero when the intruder stops, we can set eq.~\eqref{eqn:KEsol2a}
equal to zero at $z=z_{stop}$, i.e., $K_0=-\phi(z_{stop})$. The expression for $f(z)$ follows by
then differentiating with respect to $z_{stop}$, which yields:
\begin{equation}
f(z_{stop})=K_p(z_{stop})\left(\frac{dK_0}{dz_{stop}}\right)+mg.
\label{eqn:fsol}
\end{equation}
This analysis requires a determination of $K_p$ (i.e. $h(z)$ is
determined), and $z_{stop}(K_0)$, where the latter is generally
straightforward.

\section{Application to Experimental Data} 
\begin{figure}
\onefigure[scale=0.85,trim = 0mm 25mm 0mm 28mm]{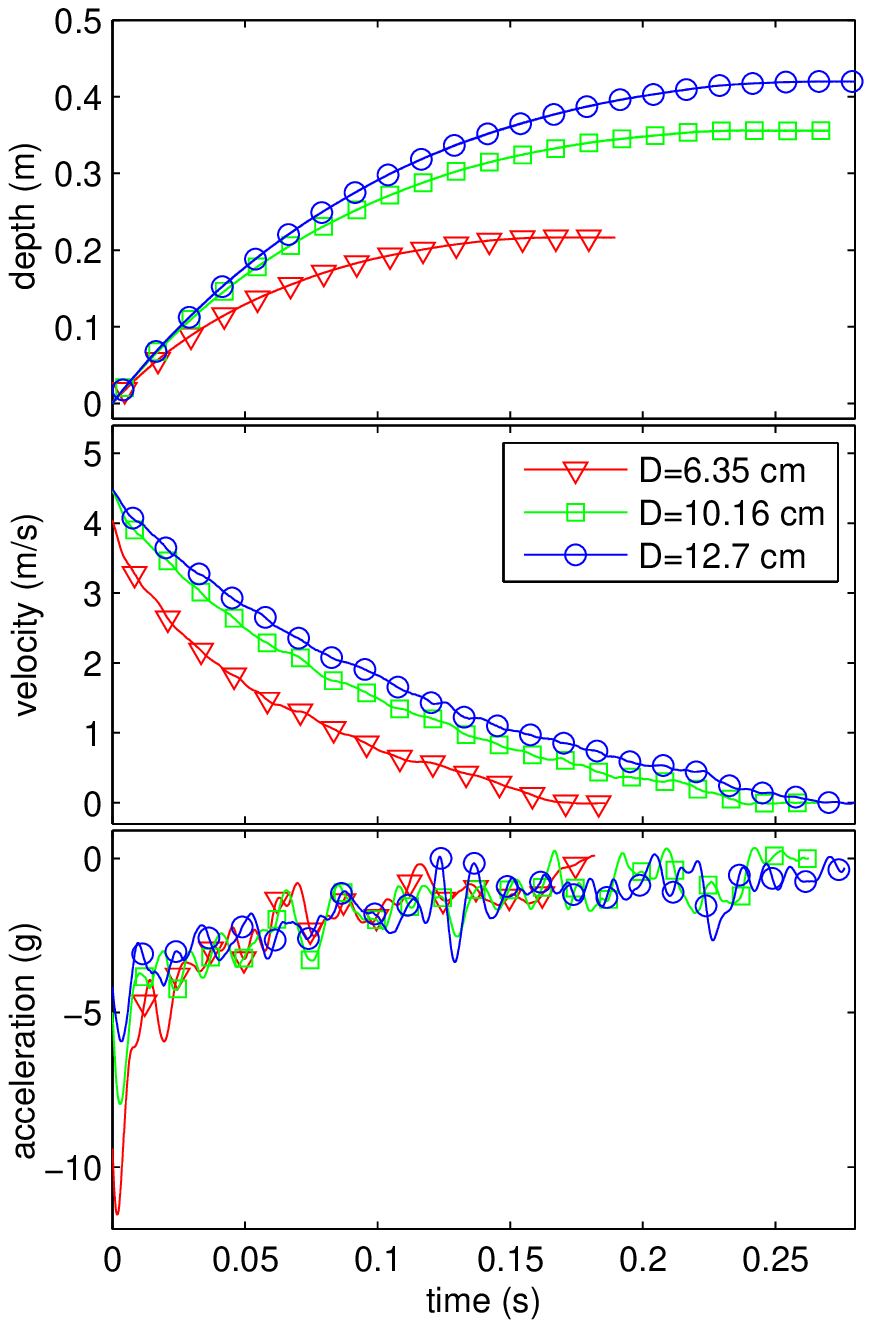}
\onefigure[scale=.35,trim=0mm 0mm 0mm 0mm]{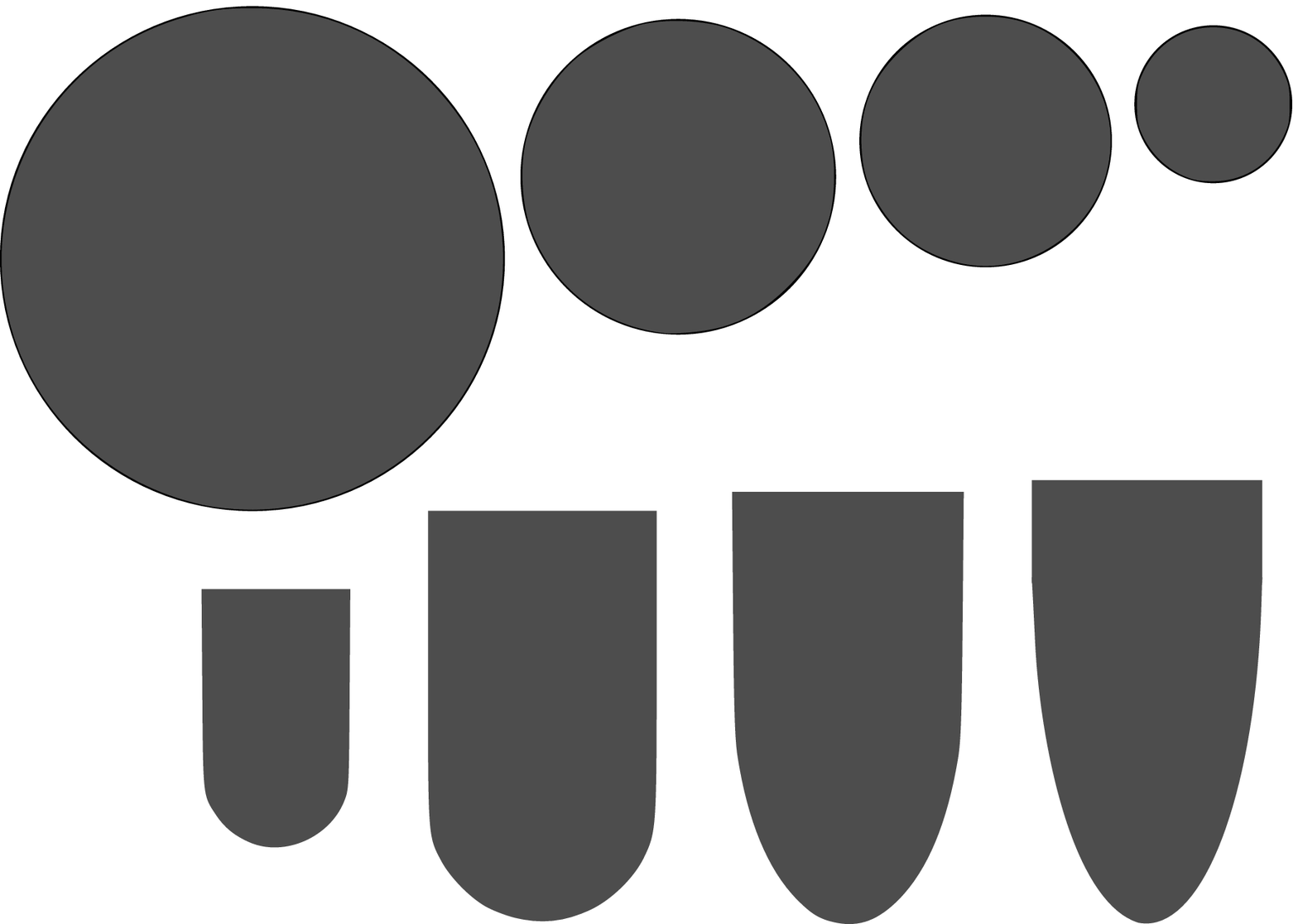}
\caption{(TOP) Single trajectories (depth, velocity, and acceleration) for the three smallest circular intruders
  with similar initial impact velocities, where $t=0$ is the first contact with the granular surface, measured
  from photoelastic response. (BOTTOM) Shapes of all intruders, drawn to scale, as described in the text.}
\label{fig:trajectories}
\end{figure}

{To test the 
approach outlined above (based on kinetic energy), and to compare to
the approach based on determining the acceleration (shown in
fig.~\ref{fig:accvsvsqr})}, we use data from two-dimensional granular
impact experiments\cite{Clark2012}, where disk and ogive intruders
(i.e. intruders with an elliptical-like nose), cut from bronze sheet,
are normally incident on a collection of approximately 25,000
bidisperse, hard, photoelastic disks (diameters of 6~mm and 4.3~mm)
confined between two 0.91~m $\times$ 1.22~m $\times$ 1.25~cm acrylic
sheets. Particles are constructed from PSM-1, a stiff photoelastic
polymer, made by Vishay Precision Group (bulk density of
1.28 g/cm$^3$, elastic modulus of 2.5 GPa, and Poisson
ratio of 0.38). Before each impact, a long rod is used to stir the particles and smooth out the top surface,
producing a fairly consistent packing fraction, $\phi\approx 0.82$. During impact,
some local compaction occurs beneath the intruder, but the global packing fraction 
does not change significantly ($\Delta\phi<0.005$). Intruders have initial
velocities between 0 and 6 m/s, which are in the subsonic regime,
since videos show the granular sound speed to be about 300
m/s\cite{Clark2012}. We record results with a high-speed camera,
typically at 40,000 frames per second, which allows us to determine
the position of the intruder in each frame, as well as the forces on
the photoelastic particles. Here, we focus on the intruder dynamics
only; sample trajectories are shown in fig.~\ref{fig:trajectories}.

Once the intruder trajectory is known, we differentiate to find the velocity and
acceleration of the intruder at each frame. To avoid noise amplification, some filtering is required
with each derivative. This is accomplished by fitting a linear
function to $W$ frames of the position data, centered at $z(t)$ for
the frame of interest, which yields the velocity, with time resolution
reduced by a factor of $W$. The same procedure is repeated to obtain
the acceleration from velocity data. We choose $W=300$, which is the
smallest value for $W$ such that the signal-to-noise ratio is 10:1 for
acceleration data. This
ensures that the observed fluctuations in velocity and acceleration
are physical. As discussed in Clark et al.\cite{Clark2012}, the
acceleration data for the intruder, obtained in this manner, exhibits
fluctuations that are intrinsic to the emission of acoustic pulses at
the interface between the intruder and grains.

Here we use eight different intruders, with width $D$ and varying nose shape,
as shown in fig.~\ref{fig:trajectories}. Four circular intruders,
with diameters, $D$, of 6.35, 10.16, 12.70, and 20.32~cm, were used to
test size effects, and four ogive intruders were
constructed to test shape effects. The shapes of the ogives consisted of a
continuous piece of material, where the leading part is a half-ellipse
truncated along the minor axis, with semi-major axis $a$ and
semi-minor axis $b=D/2$, terminated by a rectangular tail of length
$L$. Three different ellipses were used, with $a/b=1$ (half-circle),
$a/b=2$, and $a/b=3$. The width of the ogives was held constant, $D=9.3$~cm, and $L$ was varied to
keep the intruder mass constant ($L=b=4.15$~cm for $a/b=3$ case,
longer for other ogives). By keeping the width and mass constant, we
isolate shape effects. Additionally, we used one smaller
ogive with $a/b=1$, $b=3$~cm, and $L=7.7$~cm.

\begin{figure}
\onefigure[scale=0.75,trim=0mm 0mm 0mm 0mm]{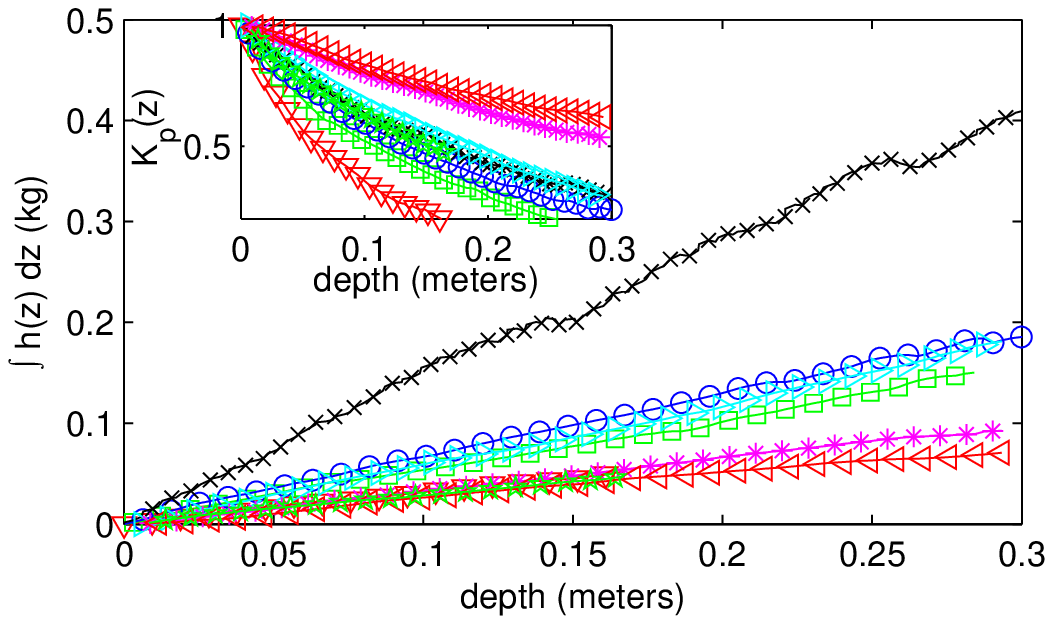}
\onefigure[scale=0.75,trim=0mm 0mm 0mm 0mm]{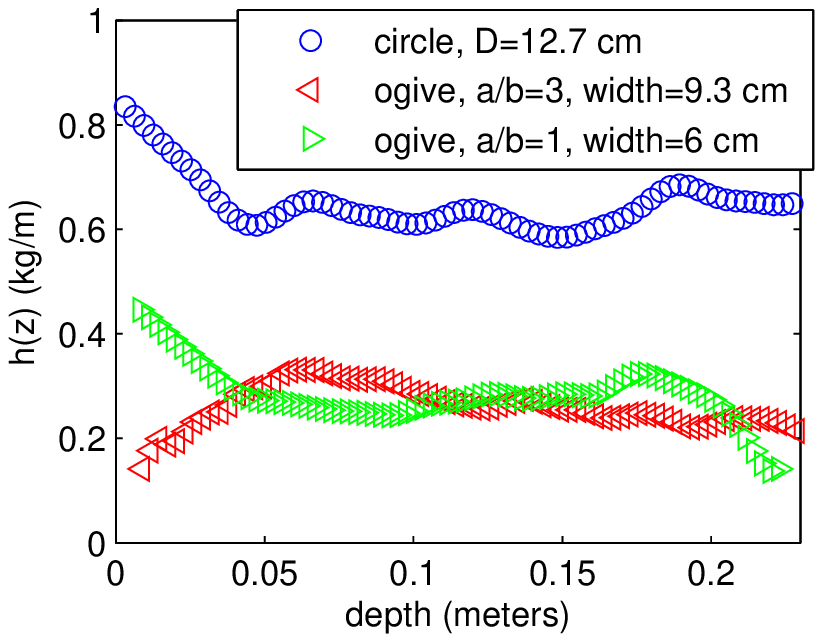}
\caption{(TOP) Plot of $\int h(z)dz =-(m/2) \log K_p$ for all
  eight intruders. All are approximately linear, and the local slope
  gives the value of $h(z)$. Inset shows $K_p(z)=\Delta K(z)/\Delta K_0$, computed
  for all pairs of trajectories, $i$ and $j$, with a minimum $\Delta
  K_0 = K_{i0}-K_{j0}$. (BOTTOM) Taking the local derivative
  shows transient behavior, which depends on the shape of the
  intruder. All circular-nosed intruders show stronger $h(z)$ 
  near the surface, reminiscent of similar effects in fluid impacts 
  \cite{Greenhow1987,Miloh1991,Mouchacca1996}.
  For the ogive with $a/b=3$, this effect is reversed: $h(z)$ starts lower and increases
  sharply to a local maximum, then weakly decreases.}
\label{fig:Kpplots}
\end{figure}

As discussed above, fitting to the force law of
eq.~\eqref{eqn:forcelaw} requires plotting the acceleration versus
velocity squared. {The fluctuations in acceleration, shown in 
fig.~\ref{fig:trajectories}, are very large, so determining a clear
value for $f(z)$ and $h(z)$ at each depth is difficult, as shown in fig.~\ref{fig:accvsvsqr}}. However, with
the kinetic energy approach, only velocity data is needed to determine
$f(z)$ and $h(z)$. Using eq.~\eqref{eqn:ijdiff}, and averaging over
all pairs of trajectories (omitting trajectory pairs with very similar
initial velocities), we obtain a clear average value for $K_p(z)$, and
thus for $h(z)$, as shown in fig.~\ref{fig:Kpplots}.

Data for $-\frac{m}{2}\log K_p = \int h(z)dz$ are approximately linear in $z$.
The slope gives the collisional term $h(z)$ (shown in fig.~\ref{fig:Kpplots}), which scales
approximately with the intruder size, as discussed below. {Here, calculating $h(z)$
requires taking a spatial derivative, which amplifies the fluctuations in $K_p(z)$. However, especially
when comparing data from multiple intruders, we are now able to separate these fluctuations 
from systematic variation in the functional form of} of $h(z)$: we observe an initial transient
regime, after which $h(z)$ approaches a nearly constant value. 
For circles, the collisional term is stronger right at impact,
which may be surprising, since the area of impact is smallest
then. This effect is reminiscent of surface tension or so-called 
``added mass'' effects for fluid impacts \cite{Greenhow1987,Miloh1991,Mouchacca1996},
but it is not obvious what physical mechanisms are at play in the
granular case. {For intruders with more elongated noses, this effect is greatly weakened, and even
reversed in the $a/b=3$ case shown in fig.~\ref{fig:Kpplots}. We also note that the nature of $h(z)$---constant, after
an initial transient---supports
the omission of as force law term which is linear in the velocity
(i.e., proportional to $K^{1/2}$), at least for the experiments discussed here.}

\begin{figure}
\onefigure[scale=.7]{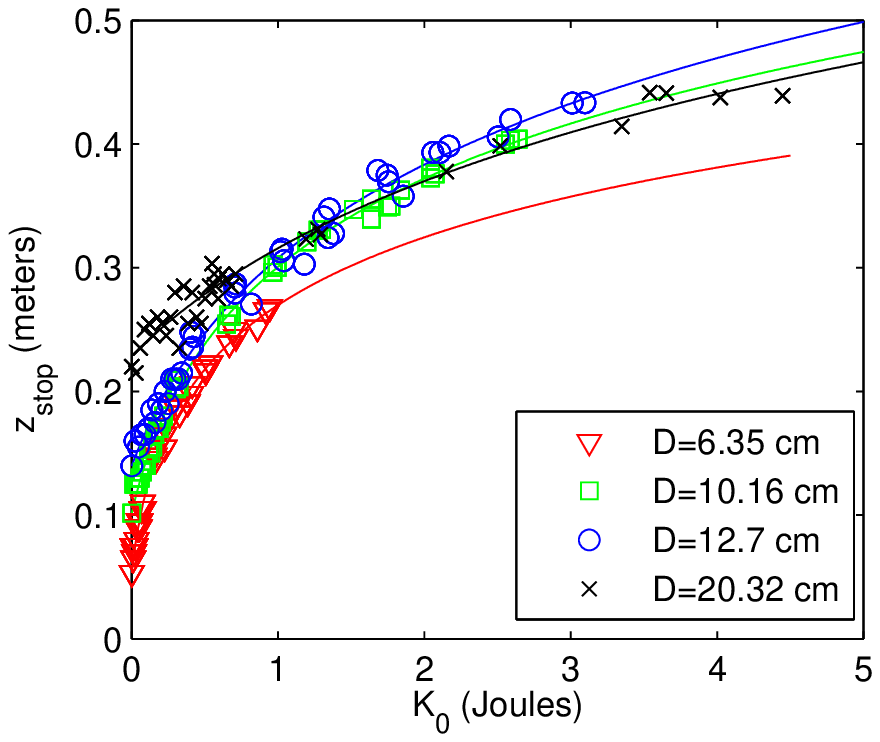}
\onefigure[scale=0.7,trim=0mm 0mm 0mm 0mm]{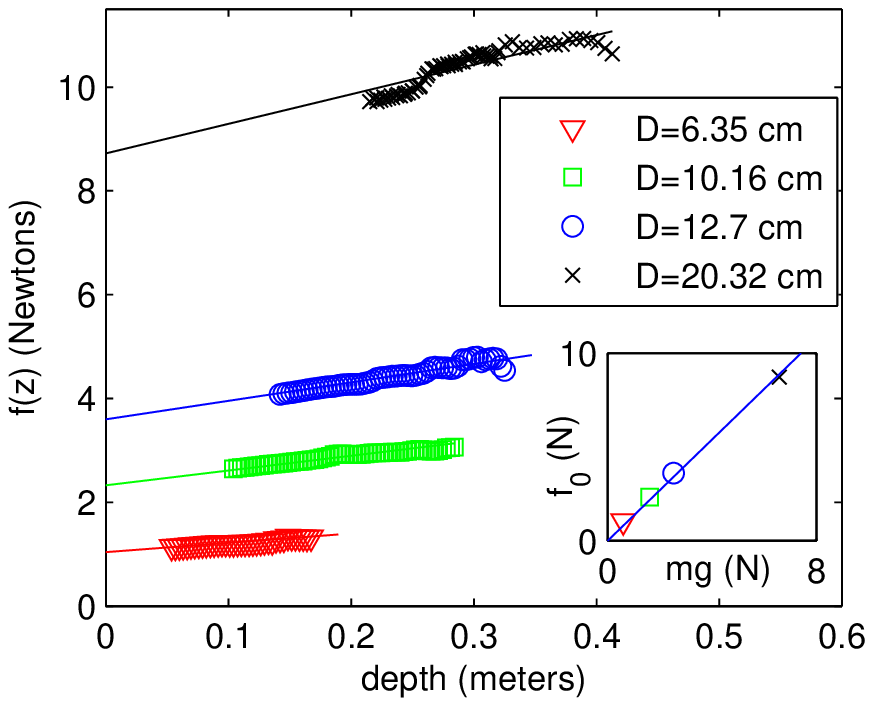}
\caption{(TOP) Plot of $z_{stop}$ vs. $K_0$, with fits of the form
  $a\log (bK_0+1)+c$. (BOTTOM) Plot of $f(z)$ for circular
  intruders. Linear fits are $f_0+kz$, where the slope, $k$,
  corresponds to hydrostatic pressure. Note that $f(z)$ is dominated
  by the offset, $f_0$, for our data. Inset shows plot of $f_0$
  vs. $mg$, with a linear fit through the origin, with slope 1.35.}
\label{fig:fzplots}
\end{figure}

Once $h(z)$ is specified, eq.~\eqref{eqn:fsol} provides a solution for
$f(z)$ with adequate data for $z_{stop}(K_0)$. First, we plot the
final depth, $z_{stop}$, versus $K_0$, as shown in fig.~\ref{fig:fzplots}. Note that the data
for the higher energies are consistent with the logarithmic behavior
in eq.~\eqref{eqn:highKEsol}, and that the data for low energies
deviate from this curve with a non-zero intercept which scales with
intruder diameter ($z_{stop}(K_0=0)\approx D$).
Only circular intruders are used, since ogive intruders penetrate much
deeper and may interact with the lower boundary of the
experiment.

Finally, by fitting a smooth function to the curve for each circular
intruder and differentiating, we solve for $f(z)$, wherever
$z_{stop}(K_0)$ is defined. This yields a linear function for all
circular intruders with a non-zero intercept, $f_0$, which scales
linearly with the intruder mass (fig.~\ref{fig:fzplots}), {i.e., $f_0\simeq 1.35\,mg$. We note that
impact experiments performed by Goldman and
Umbanhowar\cite{Goldman2008} show a similar result for $f_0$, but with a
$f_0$ which increases from 0 to $2mg$ during $0<z<D$, then saturates at
$f_0\simeq 2mg$.}

\begin{figure}
\onefigure[scale=.8, trim=0mm 0mm 0mm 0mm]{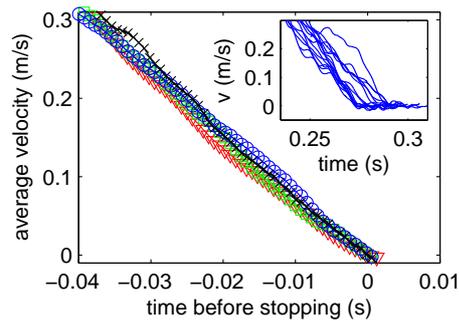}
\caption{Main plot shows the average velocity of all four circular intruders as they come to a stop. All intruders decelerate at the same rate, $a\approx 0.75 g$. The upward force required would be approximately $1.75 mg$, which is consistent with $f(z_{stop})=1.35 mg + k z_{stop}$, from fig.~\ref{fig:fzplots}. Inset shows the end of all trajectories, with initial velocities between 1 and 6 m/s, for the $D=12.7$~cm circular intruder. Note that the stopping time is very weakly dependent on initial velocity, since these trajectories all end at approximately the same time.}
\label{fig:vstop}
\end{figure}

As discussed previously, the net force must go to zero as the intruder
stops. With this in mind, we examine trajectories for $v<0.3$~m/s
(shown in fig.~\ref{fig:vstop}), where $h(z)v^2 \ll mg$.  This shows
approximately constant deceleration as the intruder comes to a stop
(consistent with the expected value of $f(z)$ from
fig.~\ref{fig:fzplots}), and the acceleration jumps to zero at $v=0$,
as in \cite{Goldman2008,Goldman2010}.

\begin{figure}
\onefigure[scale=0.75,trim=0mm 0mm 0mm 0mm]{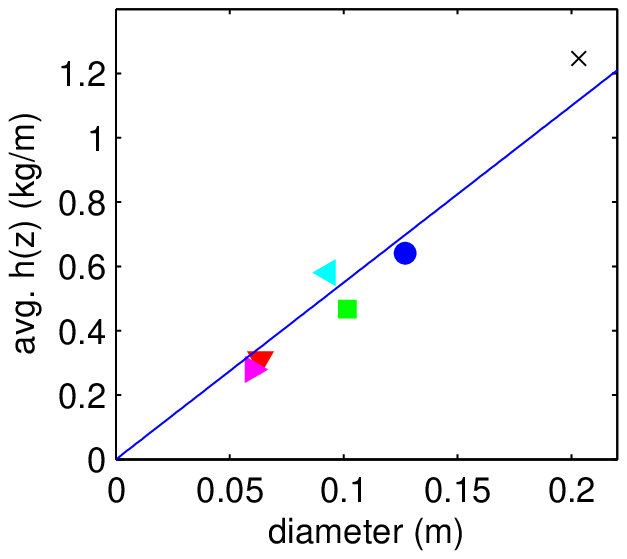}
\onefigure[scale=0.78,trim=0mm 0mm 0mm 0mm]{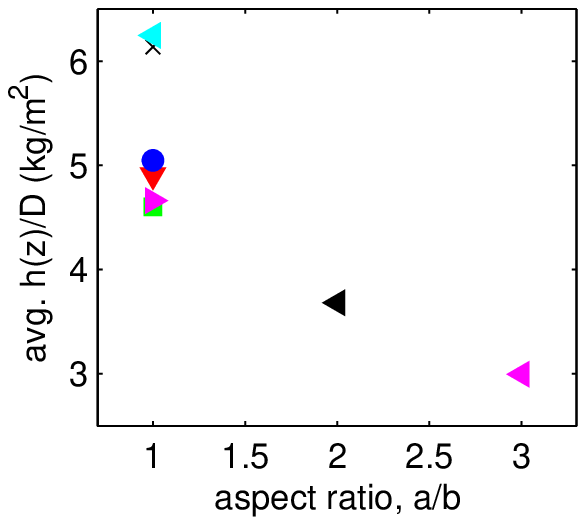}
\caption{{(TOP) Plot of the average $h(z)$ for circular-nosed intruders versus the diameter of the nose, $D$, which shows that the two are directly proportional, with $h(z)\sim 5.5D$. (BOTTOM) Plot of the average $h(z)/D$ versus the intruder aspect ratio, $a/b$, which shows a substantial decrease in the collisional force as the intruder nose is elongated.}}
\label{fig:hzvsshape}
\end{figure}

{ To examine size and shape effects, we also plot the depth-averaged
 $h(z)$ as a function of intruder size and shape,
  as shown in fig.~\ref{fig:hzvsshape}. The top plot shows that $h(z)$
  is directly proportional to the intruder size. The bottom plot shows the average $h(z)$ versus the
  aspect ratio of the elliptical nose, which falls off substantially
  as aspect ratio is increased. Thus, for equal intruder widths, a
  more elongated nose has a substantial effect in decreasing the
  collisional force. We also note that the two circular nosed
  intruders which have a slightly larger collisional force are the
  larger, circular ogive and the largest disk. We believe that the
  deviation of these two intruders relates to their larger
  mass-to-width ratio, $m/D\simeq 3$~kg/m (for all other
  circular-nosed intruders, $m/D<2$~kg/m).  These two intruders
  generate photoelastic activity that extends considerably farther
  into the material (the larger, elliptical-nosed intruders with
  $m/D\simeq 3$~kg/m generate far less photoelastic activity, perhaps
  keeping them in the same regime as the smaller circles). Thus, as a
  potential explanation, we suggest that these intruders are
  effectively interacting with a larger mass of grains.  Similar
  collective effects were used in \cite{Takehara2010}, as well as
  suggested by Waitukaitis and Jaeger to explain the more extreme case
  of shear thickening of suspensions which are subjected to impact
  \cite{Wait2012}.  }

\section{Conclusion} We have shown a new approach to
a commonly used model for describing the dynamics of an
intruder impacting on a granular material. By reformulating the model
into a linear ODE, we obtain formal solutions of the trajectory in terms of
the initial kinetic energy, as well as a systematic way of
calculating the different terms in the model---namely $f(z)$ and
$h(z)$---using only position and velocity data, which are more easily
obtained experimentally than data for acceleration.

Additionally, we have used this approach to measure $f(z)$ and $h(z)$
for experimental data. {The high level of agreement between the experimental 
data and the model in eq.~\eqref{eqn:forcelaw}, as well as the sensible behavior and scaling of the
$f(z)$ and $h(z)$ terms, validate the use of the model. However, the grain-scale origins of the force-law
terms are not well understood, and this will be the subject of future study.} We also
observe that the usual assumption that $h(z)$ is constant applies only
after an initial transient at impact, which varies with intruder
shape. This result could be important in engineering and control
applications. We also note that terms linear in velocity have been
proposed in the context of this model\cite{Goldman2008}, but our data
shows no need to implement them here.  {We also note a substantial
reduction in the collisional term, $h(z)$, as the intruder nose is elongated. Thus,} future work should include
investigation of other intruder shapes (e.g. triangular/conical
noses). {Finally, it is not clear under what conditions the force-law model is valid. Further study
might explore the limits of this model, such as intruder velocities approaching the granular sound speed
or connection to the slow drag regime \cite{Albert1999,Zhou2004,Geng2005}.}

\acknowledgments This work was supported by DTRA grant HDTRA1-10-0021, by NSF grants NSF-DMR0906908 and NSF-0835742, and by ARO grant W911NF-1-11-0110. We appreciate valuable input from Profs. Lou Kondic, Wolfgang Losert, and Corey O'Hern.

\end{document}